# System-Model-based Simulation of UML Models

María Victoria Cengarle[1], Juergen Dingel[2,3],
Hans Grönniger[3], and Bernhard Rumpe[3]

[1] Institut für Informatik, Technische Universität München, Germany
[2] Applied Formal Methods Group, Queen's University, Canada
[3] Institut für Software Systems Engineering, Technische Universität Braunschweig, Germany

**Abstract.** Previous work has presented our ongoing efforts to define a "reference semantics" for the UML, that is, a mathematically defined system model that is envisaged to cover all of the UML eventually, and that also carefully avoids the introduction of any unwarranted restrictions or biases. Due to the use of underspecification, the system model is not executable. This paper shows how the system model can serve as the basis for a highly customizable execution and simulation environment for the UML. The design and implementation of a prototype of such an environment is described and its use for the experimentation with different semantic variation points is illustrated.

## 1 Introduction

Modeling is a major activity in any development of a complex system. In contrast to all other engineering disciplines, software engineering still suffers from a distinct lack of maturity. Its foundations are still in development and discussion. The UML [1] is still undergoing heavy changes and criticism and we can expect that it will take a while until we have settled and generally agreed upon our foundations for modeling and specifying software intensive systems.

The modeling of software can serve quite a number of purposes within a development project. Commonly, models are used as high-level programming languages for generating code. Code generators and other tools for the UML do not necessarily agree on the generated structural and behavioral semantics, impeding interoperability and making the results of code generation, analysis or simulation tool dependent. This problem is rooted in one of the common criticisms of the UML, namely its lack of a precisely defined, commonly agreed semantics.

Work in [2] describes an ongoing effort to solve this problem by providing such a semantics in form of a mathematically defined *system model*, a precisely defined syntax of the UML and most importantly an explicit and unambiguously defined semantics mapping from UML syntax into the system model. This system model is defined using standard mathematical techniques while carefully avoiding any restrictions or biases that are not warranted by the UML standard.

This paper describes our work on using the system model as the basis for an execution and simulation environment. Encoding a declarative mathematical specification into an executable model requires design decisions. While some of these decisions are uncritical, such as the encoding of the universe of object identifiers as integers and variable names as strings, other decisions are more significant such as the realization of associations and the scheduling of object computation. The UML2 standard identifies many of these significant design decisions as *semantic variation points*. However, due to the informal nature of the UML specification, it is likely that the set of variation points is not comprehensive. Moreover, it is unclear what exactly good or bad choices (or values) for these variation points are, and if some of these choices interact with each other in negative or unexpected ways. While the UML2 standard sketches properties of a model of computation (see, e.g., [1, Section 6.3]) for UML models, the precise shape of that model together with its strengths and weaknesses does not become clear. The main goals of our execution and simulation environment are thus to allow us to

1. validate parts of the system model and to improve it,
2. gain a better understanding of exactly what kind of information is needed to make a UML model executable and how that information can be provided to an execution and simulation environment in a modular way that supports customization and experimentation,
3. experiment with different choices for semantic variation points, to determine "good" and "bad" choices, and uncover possible negative interactions and exclusions,
4. gain more experience with the model of computation envisioned in the UML2 standard.

While the prototype described in this paper does not achieve all of these goals yet, the results obtained are encouraging.

In the rest of the paper, we will briefly review the most relevant concepts of the system model as defined in [3–5] in Sect. 2. Then, the most interesting aspects of the design and implementation of the execution and simulation environment are described in Sect. 3. Finally, the use of the environment is illustrated in Sect. 4 by executing a given model under different scheduling strategies and observing the resulting differences in behavior. Sect. 5 sketches related work and concludes the paper.

## 2 Overview of the Theory of System Models

The goal of system models is to provide a solid mathematical basis for the definition of formal semantics of modeling languages, in particular for the UML. A system model describes the universe (set) of all possible semantic structures, each with its own behavior, of the system defined by a sentence of the syntax. Such a sentence is, in the case of UML, a complete UML specification.

Roughly speaking, a system model is a timed state transition systems, whose states embody the appropriate data structures and whose transitions mirror the behavior as defined by the sentence of interest. In order to define these kind of timed state transition systems, we made use of a combination of theories that deal with data, objects, classes, control, messages, calls, returns, recursive invocation, events, threads, time, scheduling, among other concepts.

The current state of the realization presented in Sect. 3 slightly simplifies the definition reported in [3–5]. The summary of the present section, thus, only addresses the relevant parts of the system model definition as used in the next section.

Different principles guided the definition of the theory of system models. Instead of using a specialized notation such as, e.g., Z, B, or ASMs, we relied on mathematics directly, in particular on the theories of numbers, sets, relations, and functions. Those other notations are biased towards certain purposes as for instance model checking analysis and thus are less versatile.

The theory of system models does not constructively define its constituents, i.e., its elements are descriptively characterized. Moreover, no implicit assumptions are made: what is not explicitly specified, need not hold. In addition, the theory of system models presents various so-called semantic variation points, that allow a specialization of the theory to particular realms. Other principles that guided the definition of system models can be found in [3–5].

A system model is a timed state transition system. States of a system model are composed by static information (basically values and store), dynamic information (basically threads and control), and a finite number of message pools (in the simplest case, queues of incoming messages like method invocation and signals). A data store is a partial function

$$\mathsf{DataStore} : \mathsf{UOID} \hookrightarrow \mathsf{UVAL}$$

that maps object identifiers onto values. DATASTORE denotes the universe of all possible data stores. CONTROLSTORE denotes the universe of all possible control states. A control state contains information about all the threads being executed:

$$\mathsf{ControlStore} : \mathsf{UOID} \hookrightarrow \mathsf{UTHREAD} \hookrightarrow Stack(\mathsf{UFRAME}),$$

where the universe of execution frames is defined by

$$\mathsf{UFRAME} = \mathsf{UOID} \times \mathsf{UOPN} \times \mathsf{UVAL} \times \mathsf{UVAL} \times \mathsf{UPC} \times \mathsf{UOID}.$$

That is, a frame contains the identifier of the executing object, the operation invoked, parameter values packed in a record, values of the local variables likewise packed in a record, the program counter, and the identifier of the caller object. Finally, UEVENT denotes the universe of possible events. The events of an object are retrieved via the function

$$\mathsf{EventStore} : \mathsf{UOID} \to \wp(\mathsf{UEVENT}).$$

The universe of all possible event stores is denoted by EVENTSTORE.

States of a system model are thus defined as triples $(ds, cs, es)$, where $ds \in$ DATASTORE, $cs \in$ CONTROLSTORE, and $es \in$ EVENTSTORE. How the three stores are used to capture the state of a single object and the system is explained in more detail in Sect. 3.

A *system model* is a tuple (STATE, $\Delta$, Input, Output, Init) with

- STATE a set of states of the form $(ds, cs, es)$ as described above,
- Input and Output the input and output channel sets, respectively,
- Init $\subseteq$ STATE, and
- $\Delta : (\text{STATE} \times \text{T}(\text{Input})) \rightarrow \wp(\text{STATE} \times \text{T}(\text{Output}))$ a transition function

where $\text{T}(C)$ is the set of possible channel traces for the channel set $C$, i.e., if $x \in \text{T}(C)$ and $c \in C$, then by $x.c$ we denote the finite sequence of messages $x(c) \in \text{UMSG}^*$ where UMSG is the universe of all messages.

Timed state transition systems behave like a Moore machine,[4] that is, the output depends on the state and not on the input. Moreover, the state transition function is total. The theory of system model encompasses further definitions, like composition of system models and interface abstraction. These provide comfortable tools that allow compositional definition and abstraction from representation details. They moreover fulfill interesting and desirable properties. Due to lack of space, the interested reader is referred to [3–6].

Composition allows to independently implement parts of a specification and subsequently compose these; the system model this way obtained implements the complete UML specification. The virtual machine presented in Sect. 3 below implements the state transition function. This implementation does not reflect this abstract view on the composed system directly. Instead, a fine-grained account on control flow and the effect of actions on an object's state is given.

## 3 Design and Implementation of a System Model Simulator

This section provides an overview of the current state of an implementation of the system model in Haskell. A basic knowledge of Haskell is assumed. For more information see [7, 8].

The current state of the implementation differs from the system model definitions given in Sect. 2 and in [3–5] in various ways: Some simplifications have been made that will, however, be brought in line with the mathematical definitions as the implementation advances. Furthermore, the mathematical definitions allowed us to leave quite a number of execution relevant issues unspecified that have to be considered in order to make the system model executable. In general, the predicative specification style is replaced by an executable simulator that allows for experimentation with different choices for variation points. Just like in the declarative system model, we tried to avoid unnecessary overspecification. Whenever design decisions were necessary to achieve executability, we implemented these in a modular, replaceable way. We, e.g., had to provide constructive solutions for scheduling or an action language. In developing an action language we tried to stay in line with the UML standard in the sense "that all behavior in a modeled system is ultimately caused by actions", cf. [1, Sect. 6.3.1].

---
[4] A Moore machine is a finite state machine that produces an output for each state.

An important decision to make is, which language to use for implementing the simulation engine. We decided against the natural choice of using an object-oriented language, but use the lazy functional language Haskell for several reasons. First, Haskell is more similar to mathematics and therefore allows us to stay closer to the mathematical definitions. Second, with Haskell very compact specifications can be defined. And third, constant confusion arises if the encoding language is conceptually close to the encoded language, i.e., it is always yet to clarify whether it is a simulated class or a class used to handle the simulation. Haskell is conceptually distant enough from the UML so that such problems do not occur. Furthermore, this conceptual distance enforces a deep embedding of the UML into Haskell. For example the typing system of the UML is independent from the Haskell type system and is completely encoded within a universe of types on its own.

In Sect. 3.1 we first explain the architecture of the system model simulator and highlight the main concepts necessary for this paper from each module. Data needed to set up the execution and the several execution steps themselves are described in Sect. 3.2. Finally, the currently realized choices for variation points of the system model are discussed in Sect. 3.3.

### 3.1 Architecture

The architecture of the implementation has two conceptual layers. First, the system model is implemented by a set of Haskell modules and provided to the simulator by a dedicated module *SystemModel*. The system model part of the architecture again is built in a layered form where sophisticated concepts depend on more fundamental definitions. The system model modules mirror the definitions from the mathematical version but are extended to incorporate functions needed for executability. Fig. 1 shows the architecture as a set of dependent Haskell modules. Modules *Map*, *Stack*, and *Buffer* are general purpose implementations with the obvious operations.

**Basics** *StructureBasics* defines structure-related system model entities. The universe of type names, for example, is defined as a data type

```
UTYPE = TInt | TVoid | .. | XClass UCLASS
```

where `UCLASS` is a class definition. "Primitive" types as `TInt` start with a `T`. Because `UCLASS` (universe of class names) mathematically is a subset of `UTYPE` but defined on its own, it is wrapped by the type constructor `XClass` to become part of `UTYPE`. The universe of values `UVAL` introduces type constructors which start with a `V`, e.g. `VInt 2` represents the integer value 2 and `VVoid` the singleton value "void". The universe of object identifiers `UOID` is integrated into `UVAL` by introducing the type constructor `XOID`. Further, this module provides a definition of a subclassing relation as `SubclassRel = Map UCLASS [UCLASS]` modeling multiple inheritance, if needed. `UOPN` defines the signature of a method and is defined in *StructureBasics*, while `UMETH`, which among other things contains a list of actions that constitute the method's body, is defined in *BehaviorBasics*. The mapping `MethMap` records for each class C the operations contained in C and which methods implement each of these operations. Module *BehaviorBasics* also contains definitions of `UFRAME` and `UTHREAD` which roughly coincide with the mathematical definitions.

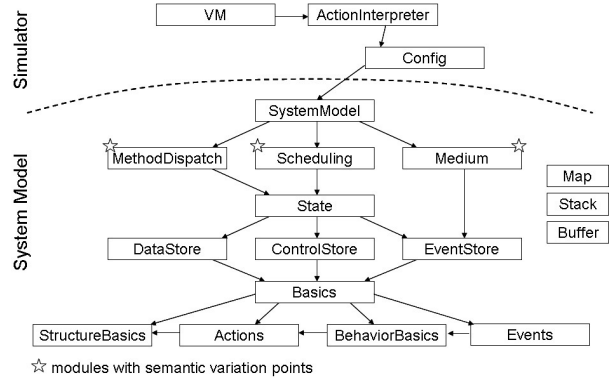

**Fig. 1.** Architecture of the system model in Haskell

Module *Actions* contains a data type that defines the action language that is interpreted by the *ActionInterpreter*. It contains actions for sending messages, creating objects, setting attributes or variables and for doing basic arithmetic. As an example, actions will be used in Sect. 4 to provide an implementation for a method. Events defined in module *Events* are `CallEvent`, `ReturnEvent` and `SignalEvent` each associated with a message of type `UMSG`. The events are used to trigger synchronous operation invocation, operation return or asynchronous communication, respectively. For convenience, the list of basic modules is collected in a module *Basics*.

**State** The state of the system model (module *State*) is determined by the three stores, data, control, and event store, implemented in the respective modules. So, the state is of type `USTATE = (DataStore, ControlStore, EventStore)`. Stores are equipped with functions to retrieve the current state of an object (attribute values, currently executing threads, and pending events). Additionally, functions are provided to construct empty stores, or to add or update state information for a given object.

**SystemModel & Simulator** All interaction between objects (asynchronous communication as well as method invocation) is achieved by the generation of events that lead to messages passed from one object to the event buffer of another. The module *Medium* provides a function that "transmits" messages between (potentially physically) distributed objects over some communication media. The concrete implementation of the medium (e.g. lossy or reliable transmission) is a semantic variation point (cf. Sect. 3.3) and can easily be replaced in the simulator. Another variation point is given by the way how method calls are resolved. Various approaches exist (e.g. single or multiple dispatch) and can all be used by implementing a method `MethodDispatcher` in module *MethodDispatch* appropriately.

The way computation is scheduled in the system model is largely left open in the mathematical definitions. In the implementation, the required function signatures have been fixed in the module *Scheduling*. Concrete scheduling strategies can be realized as a further variation point.

The described modules are provided to the simulator through the module *SystemModel*. The simulator consists of three modules. *ActionInterpreter* interprets each basic action while the virtual machine *VM* schedules the next object and thread for execution based on the system model configuration *Config*.

### 3.2 Execution

This section provides a more detailed description of the basic steps that constitute a simulator run.

**Input to Simulator** In order to run the simulator, it has to be provided with three kinds of data:

1. Decisions on specific variation points in the system model (e.g., decisions about the kind of scheduling, method dispatching etc.).
2. Description of the system to be simulated in terms of the system model (definitions of classes, methods etc.).
3. Initial setup, that is, the state of the system in which execution starts (generally, a network of initially created objects that may start executing an operation).

The selection of variation points and the system description will be used throughout the simulation and is handed over to the simulator in form of a data type `Config` together with the following functions

- `dispatcherOf :: Config -> MethodDispatcher`
- `schedulerOf :: Config -> (RunnablesSel, Scheduler)`
- `mediumOf :: Config -> Medium`
- `subclassRelOf :: Config -> SubclassRel`
- `methMapOf :: Config -> MethMap`

The first three functions return the specific choices of the user for the different variation points concerning method dispatching, scheduling, and the type of medium.

The specification of the simulated system is typically provided in a separate Haskell module that contains class and method definitions, and a definition of `Config`. Additionally, the initial setup may be specified as a type

```
Setup = [(String, UCLASS, OKind, [String])]
```

In the setup it is decided whether an initial object should be active or passive. The data type `OKind = Active UOPN Prio | Passive` either contains an operation and a priority or it indicates that the object should be passive. The setup is used as input for function `runMain` of the virtual machine that constructs the initial state. For each entry in the setup list `(a,cl,k,cons)` an object of type `cl` is created, either starting execution in operation `op` with priority `prio` if `k` matches the pattern `Active op prio` or remaining passive, and having a link to each of the other objects in list `cons`. The objects at this point are identified via a string `a` because their actual object identifier values are not yet known. Usage of `runMain` is optional as the initial state may also be created manually in order to start the simulator via the function `run`.

**Top-level loop of VM** Module *VM* implements the following function signatures to start execution:

```
runMain :: (Config, Setup) -> (USTATE, Time)
run :: ((Map ThreadID Time), Time, Config, USTATE) -> (USTATE, Time)
```

Function `runMain` constructs the initial state (of type `USTATE`) and calls `run`. After the simulation is completed the final state and time is returned. The function `run` forms the "main loop" of the simulation and is called recursively at the end of each step. It takes two additional arguments. The first parameter is a map that stores the last execution time for each thread and the second provides a notion of the current execution time (which counts the number of simulation steps). The Haskell code in Fig. 2 summarizes the behavior of `run`.

```
1 run (times, t, conf, state) =
2   let (rselector, schedule) = schedulerOf (conf)
3       runnables = collectRunnables (rselector, state)
4       runnables' = addLastExecInfo (times, runnables)
5       (oid, thread) = schedule (t, runnables')
6       state' = exec (oid, thread, state, conf)
7       times' = update (times, (thread, t))
8   in if (null runnables) then (state, t)
9      else run (times', (t+1), conf, state')
```

**Fig. 2.** Haskell code for function `run`

The process of scheduling the next object and thread is done in two steps. First, given an object and a state, a function `rselector` collects the threads of the object that are ready for execution. In line 2, the type of `rselector` is

```
RunnablesSel = (USTATE, UOID) -> [(ThreadID, Prio)]
```

For each object the information available to decide which threads to offer for scheduling is the whole state of the object. In particular, this function is able to inspect the object's event buffer to check whether newly arrived events need to be processed. The function called in line 3 is a higher-order function defined in module *Scheduling* of signature

```
collectRunnables :: (RunnablesSel, USTATE) -> [(UOID, ThreadID, Prio)]
```

that collects the information about runnable threads for each existing object in the data store using the function given as the first argument. If the list of runnables is empty, the execution ends and the final state is returned (line 8). Otherwise, the list is extended with the last execution times of each thread after which it is of type `[(UOID, ThreadID, Prio, Time)]` (line 4). Now, the second part of the scheduling process takes place. Function `schedule` of type

```
Scheduler = (Int, [(UOID, ThreadID, Prio, Time)]) -> (UOID, ThreadID)
```

schedules the next object and thread combination.

The next atomic execution step is carried out in line 6 using the function `exec` defined in module *VM*, and will be explained with the help of the Haskell code given in Fig. 3 below. Finally, the map with the last execution times is updated and `run` is called recursively (lines 7 and 9).

```
1  exec :: (UOID, ThreadID, USTATE, Config) -> USTATE
2  exec (oid, threadid, state, conf) =
3    let state' = consumeEvent (state, conf, oid, threadid)
4        cs = csOf (state')        -- get the controlstore from state
5        thread = threadOf (cs, oid, threadid) -- get the thread
6        md = dispatcherOf (conf)
7        scl = subclassRelOf (conf)
8        mmap = methMapOf (conf)
9        ds = dsOf (state')        -- get the datastore from state
10       op = opOf (thread)        -- get the operation from thread
11       (Meth _ _ actions) = md (scl, mmap, ds, oid, op)
12       pc = pcOf (thread)        -- get the pc from thread
13       action = actions!!pc
14       state'' = interpret (action, state', thread, conf)
15   in state''
```

**Fig. 3.** Haskell code for function `exec`

**Carrying out an atomic step** The function `consumeEvent` called in line 3 of function `exec` (Fig. 3) removes an incoming event (if any) from the object's buffer and prepares the context for handling this event. In case of a method call, it constructs a thread frame containing all relevant information (parameter values, etc.); in case of a return event it restores the thread's context and provides the return value as a local variable. For an asynchronously received event a thread is created that starts executing the operation stated in the event's message. In lines 4-10, information used for method dispatching is extracted from the configuration and state. The method dispatcher `md` is a function with signature

```
MethodDispatcher = (SubclassRel, MethMap, DataStore, UOID, UPON) -> UMETH
```

Based on the system description (subclassing and relationship between classes, operations and methods), the data store (used to retrieve the type of the object), the object identifier and the operation to call, the actual method is determined and returned. The returned method contains a list of actions (line 11). The next action to execute is pointed to by the current value of the program counter of the thread. Interpretation of the action is done by calling the function `interpret` of the module *ActionInterpreter*. For each action defined in module *Actions* there exists a corresponding part of the interpreter function that is selected by the pattern matching mechanism of Haskell. The signature is

```
interpret :: (Action, USTATE, UTHREAD, Config) -> USTATE
```

In general, this function makes changes to the system's state (by altering attribute values, creating events, adjusting the program counter, etc.). If the action involves sending of an event to another object (e.g. in case of an operation call), `interpret` accesses the medium using the function `mediumOf` on the `Config`. The function `Medium = (EventStore, Event) -> EventStore` is responsible for putting the event in the right event buffer of the receiving object.

### 3.3 Variation Points

Variation points in the system model implementation mark points where different realization variants of a function may lead to different behavior or efficiency of the simulator and therefore the simulated system. To allow experimentation, the Haskell implementation of the system model provides different alternative realizations of variation points that can be exchanged in different runs of the simulator. Currently, the following variation points are considered in the implementation explicitly:

1. Scheduling
2. Method dispatching
3. Event distribution by a medium

Selecting implementation alternatives for these tasks is possible by using the data type `Config` in different combinations. In the future, other variation points mentioned in [3–5] will be added, for example, different realizations of associations in the system model which have not yet been considered in the implementation. Facilitating alternative structures or behavior is also possible by replacing one or more modules by alternative implementations. This, for example, allows for adding additional basic types to the UTYPE structure.

Up to now, we have only started to explore the scheduling variation point in greater detail. For the other two variation points single implementations are provided. Method dispatch (supporting single inheritance) is done by looking up a method in the class of the object; if not found there, the method is looked for in the superclass(es). The medium is implemented as a "reliable" medium that without loss or reordering transfers an event into the receiver's event buffer.

**Scheduling** Recall that scheduling is performed in two steps. First, a function of type `RunnablesSel` is used to retrieve a list of runnable treads for each object. Two alternative implementations are provided:

**RTC** The function does not consider any event in the buffer while another thread is still active in the object. This is also known as *run-to-completion* execution.

**CONC** As soon as an event is put in the event buffer, the function also offers the thread that would handle this event to the scheduler, leading to potentially *concurrent* execution of multiple threads in one object.

Second, based on the current list of runnable threads, one thread in one object is scheduled for execution. Here, also two variants of scheduling strategies have been implemented:

**RR** Round-robin scheduling selects all threads alternately.

**PRIO** A priority-based scheduling finds the thread with the highest priority and smallest last execution time and selects it for execution. Priorities change dynamically in that the effective priority is computed by a thread's base priority plus its waiting time (aging).

Although we provide a single function for scheduling all objects, the implementation of more complex scheduling strategies that make use of different scheduling domains and different "sub schedulers" (e.g. to handle groups of objects that live on the same processing node) is also possible.

## 4 Example

The purpose of this section is to show how a concrete system can be described in terms of the system model and how to configure and run the simulator with different values for the choices for the variation points.

The system we are going to simulate is depicted in Fig. 4. The *Producer* is an active object executing its *produce* method that produces two data elements (in this example the integer values 10 and 20 will be produced) and provides them by calling method *put* on the buffer before it terminates. We assume that producing a new data element takes a long time. This is modeled by "counting to 5" each time before making one data element available. Method *put* sets the attribute *data* to the received value, overwriting the stored value whether or not it has been fetched before. A *Consumer* is also an active object that tries to fetch exactly one data element from the buffer. It continuously calls *get* which returns -1 if no data is available. After successfully getting a value from the buffer, the consumer sets its *data* attribute to that value and terminates. The method *get* sets the buffer's attribute *data* to -1 if a consumer has fetched the current value so that no other consumer will get the same value.

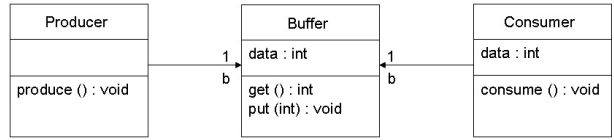

**Fig. 4.** Example system

Despite its simplicity, this example suffices to show the effects of the different variation points when running the simulator.

Due to lack of space, we only show how the class *Buffer* and its *put* method are described in terms of the system model. In Fig. 5, lines 1-2 show a class declaration (constructor `MClass`), the attribute of a class `buffer` is kept in a record (constructor `MRec`), is called `data` and is of type `TInt`. In lines 4-5 operation `put` is declared. It takes a single parameter of type `TInt` and its return type is `TVoid`. The definition of a method implementing operation `put` starts in lines 7-8 by stating that it implements operation `putOp`, that the parameter name is `p` and that the method's body is given as a list of actions called `putActions`. In line 10 a local variable of name `d` with type `TInt` and initial value zero is created. The following action assigns the value of the parameter `p` to the local variable `d`. The data attribute is set in line 12 and finally `VVoid` is returned.

```
1  buffer :: UCLASS
2  buffer = MClass "Buffer"  (MRec [("data", TInt)])
3
4  putOp :: UOPN
5  putOp = MOp "put" [TInt] TVoid
6
7  putMeth :: UMETH
8  putMeth = MMeth putOp [("p",TInt)] putActions
9
10 putActions = [CreateLocalVariableAction "d" TInt (VInt 0)
11              ,SetLocalVariableFromParamAction "d" "p"
12              ,SetAttribAction "data" "d"
13              ,SendReturnAction (VVoid)]
```

**Fig. 5.** Class and method definition for the Buffer

In the same way, the producer class `prod`, the consumer class `cons` and their operations and methods are declared. The next step in the system description is to connect classes, operations and methods. To this end, `methMap` maps class `buffer` to the map that defines that operation `putOp` is implemented by `putMeth` and `getOp` by `getMeth`, and maps classes producer and consumer to operations `prodOp`, `consOp` implemented in `prodMeth` and `consMeth` respectively. In this example we do not have subclassing so we define the subclassing relation to be the empty map `subc = emptyM`.

The initial setup in which execution should start is shown in Fig. 6. The initial system state will have one passive object of type `buffer` (identified for further reference by `"b"`) and three active objects: two consumer objects `"cons1"` and `"cons2"` execute operation `consOp` with priority 1, and having a link to `"b"`. Also, one producer `"prod1"` executing operation `prodOp` with higher priority 10 is created. With the help of a framework for domain specific languages [9], all Haskell definitions have been generated from textual versions of UML's class diagrams (with lists of actions for method implementations) and object diagrams (for the initial setup).

```
1 setup = [("b"    , buffer, Passive,          []),
2          ("cons1", cons,   Active consOp 1,  ["b"]),
3          ("cons2", cons,   Active consOp 1,  ["b"]),
4          ("prod1", prod,   Active prodOp 10, ["b"])]
```

**Fig. 6.** Initial setup for simulation

The first concrete configuration `conf` of type `Config` with which the simulator is called uses standard implementations for method dispatching and medium as discussed in Sect. 3.3. For the scheduling strategy, we'll start with concurrent execution of threads in one object (CONC) and a round-robin scheduling (RR). Calling function `runMain (conf, setup)` computes a final system state. Fig. 7 shows only the console output of the data store of the final state with the attributes and values for each object. It can be seen that the two consumer objects both received the same value "10" and the second value "20" still remains in the buffer. The concurrent execution of the method `get` thus led to inconsistencies illustrating the fact that the buffer code is not thread-safe (in a different setup other inconsistencies, e.g., due to concurrent `put` and `get` are possible, too). The total number of steps needed for simulation is 221.

For the second run we use run-to-completion execution (RTC) and priority-based thread scheduling (PRIO). The resulting data store is depicted in Fig. 8. The buffer in the final state is empty and each consumer received one of the two produced values. Since only one thread is allowed to "enter" the object `buffer`, no inconsistencies are possible. Moreover, the number of steps required is reduced to 64% compared to the previous run because the computation intensive production of data elements in the producer received a higher priority. The two

```
1 attributes:
2 Producer(id 0): [("b",XOID 3)]
3 Consumer(id 1): [("data",VInt 10)
4                 ,("b",XOID 3)]
5 Consumer(id 2): [("data",VInt 10)
6                 ,("b",XOID 3)]
7 Buffer(id 3): [("data",VInt 20)]
8 time: 221
```

```
1 attributes:
2 Producer(id 0): [("b",XOID 3)]
3 Consumer(id 1): [("data",VInt 20)
4                 ,("b",XOID 3)]
5 Consumer(id 2): [("data",VInt 10)
6                 ,("b",XOID 3)]
7 Buffer(id 3): [("data",VInt -1)]
8 time: 142
```

**Fig. 7.** Result with concurrent threads and round-robin scheduling

**Fig. 8.** Result with run-to-completion and priority-based scheduling

other combinations, (CONC + PRIO) and (RTC + RR), lead to faster execution with inconsistent results or to slower execution with correct results, respectively.

## 5 Conclusion

In this paper we described the implementation of a virtual machine for the simulation of UML-specified systems. The virtual machine does not interpret UML models directly, instead UML models can be mapped into the system model that is then simulated under consideration of the specific choices for the variation points. The development of the simulator allows us on the one hand to validate the system model given in [3–5] and on the other to experiment with different choices for variation points.

Our work differs from the work in [10] in that we also support the description of behavior while [10] focuses on structural modeling. The virtual machine described in [11] focuses on traceability of models. UML class diagrams and sequence diagrams are supported. Code generation is parameterized to some extent. In the industrial area various commercial tools exist that all implement a certain executable subset of UML with a fixed semantics, e.g. [12]. The work closest to ours is that of [13] in which a generic model execution engine is used as a basis for a UML simulator that also supports semantic variation points. Moreover, there exists a wealth of related work regarding the definition of UML semantics. Most approaches, however, focus on the semantics of one or two diagrams (e.g. [14–16]) whereas we aim at defining semantics for a larger set of diagrams. Yet, a detailed discussion of related work in that area is beyond the scope of this paper.

The development of our simulator is still ongoing. Some concepts from the mathematical definitions still have to be integrated (e.g. associations) and support for more variation points has to be added. Further, the basic actions and execution mechanisms may also be compared and aligned with the results of the OMG's standardization of Executable UML [17]. As a next step, it is planned to systematize the actual mapping from UML diagrams to the system model and to identify additional variation points in the mapping.

**Acknowledgement:** The work presented in this paper is undertaken as a part of the MODELPLEX project. MODELPLEX is a project co-funded by the European Commission under the "Information Society Technologies" Sixth Framework Programme (02- 06). Information included in this document reflects only the authors' views. The European Community is not liable for any use that may be made of the information contained herein. This research is supported by the DFG as part of the rUML project.